# Properties of Modal Quality Factors

Wen Geyi

*Abstract*—Spherical wave functions play an important role in the theoretical study of antenna. When they are used to investigate the stored energy outside the circumscribing sphere of the antenna, two different types of modal quality factors appear which exhibit some interesting properties. These properties can be easily demonstrated by numerical tabulations but have never proved rigorously and have remained unsolved for many years. An attempt is made in this paper to try to solve these longstanding problems. New properties and new power series expansions for the modal quality factors have been obtained, which essentially belong to the spherical Bessel functions and therefore are universally applicable.

## I. INTRODUCTION

The study of antenna quality factor $Q$ can be traced back to Chu's work published in 1948[1]. Chu's analysis is based on the spherical mode expansions and is valid for a special type of omni-directional antenna that only radiates either TE or TM modes. Chu used the total energies (instead of stored energies) outside of the circumscribing sphere of the antenna to study the antenna $Q$. In order to avoid the difficulty that the total electric and magnetic field energies outside the circumscribing sphere are infinite, Chu introduced the equivalent impedance for each mode and obtained an expression of antenna modal $Q$, denoted by $Q_n^{CHU}$ through the calculation of stored energies in the truncated equivalent ladder circuit for the impedance. Collin and Rothschild [2] adopted a method for evaluating antenna $Q$ in 1964, which was based on an idea proposed by Counter [3] that the stored energy could be calculated by subtracting the radiated field energy away from the total energy in the fields, and they introduced the modal quality factor $Q_n$ defined by

$$Q_n(x) = x - \left|h_n^{(2)}(x)\right|^2 \left[\frac{x^3}{2} + x(n+1)\right] - \frac{x^3}{2}\left|h_{n+1}^{(2)}(x)\right|^2$$
$$+ \frac{2n+3}{2}x^2\left[j_n(x)j_{n+1}(x) + n_n(x)n_{n+1}(x)\right], \qquad (1)$$
$$n = 1, 2, \cdots.$$

Manuscript received October 5, 2015. This work is supported in part by the Priority Academic Program Development of Jiangsu Higher Education Institutions, in part by a grant from Huawei Technologies Co. Ltd.
The author is with Research Center of Applied Electromagnetics, Nanjing University of Information Science and Technology, Nanjing, 210044, China. (e-mail: wgy@nuist.edu.cn).
.

where $x = ka$ with $k$ being the wavenumber and $a$ being the radius of the circumscribing sphere which encloses the antenna; $j_n(x)$ and $n_n(x)$ are spherical Bessel functions of the first and second kind respectively; and $h_n^{(2)}(x)$ is the spherical Hankel function of the second kind. The method used by Collin and Rothschild was further extended to study the quality factor for an arbitrary antenna in terms of spherical wave expansions by Fante in1969 [4]. In addition to the modal quality factor $Q_n$, Fante introduced another modal quality factor $Q_n'$ defined by

$$Q_n'(x) = x - \frac{x^3}{2} \cdot \left[\left|h_n^{(2)}(x)\right|^2 - j_{n-1}(x)j_{n+1}(x) - n_{n-1}(x)n_{n+1}(x)\right],$$
$$n = 1, 2, \cdots.$$
$$(2)$$

This is quite different from the modal quality factor defined by (1). By use of recurrence relations for spherical Bessel functions, it is readily found from (1) and (2) that

$$Q_n(x) - Q_n'(x)$$
$$= \left|h_n^{(2)}(x)\right|^2 nx - x^2\left[j_n(x)j_{n-1}(x) + n_n(x)n_{n-1}(x)\right] \qquad (3)$$
$$= -x(n+1)\left|h_n^{(2)}(x)\right|^2 + x^2\left[j_n(x)j_{n+1}(x) + n_n(x)n_{n+1}(x)\right].$$

Since Chu has neglected part of stored energies due to the truncation of the equivalent ladder network, the modal quality factor $Q_n(x)$ is larger than $Q_n^{CHU}$. Note that all the studies mentioned above only considered the stored energies outside the circumscribing sphere of the antenna in the calculation of antenna quality factor. For this reason, the quality factor obtained is much smaller than its real value.

Both the modal quality factor $Q_n$ and $Q_n'$ are important in the theoretical research, and they have been extensively used to study the general properties of antenna, such as the lower bund of quality factor and the upper bounds for the ratio of gain to quality factor for an arbitrary antenna [5]-[10]. Some properties of the modal quality factor $Q_n$ and $Q_n'$ have been revealed by Fante through numerical tabulations [4]:

$$Q_n'(x) > 0. \qquad (4)$$

$$\frac{Q_{n+1}'(x)}{Q_n'(x)} > 1. \qquad (5)$$

$$Q_n(x) > 0. \qquad (6)$$

$$\frac{Q_{n+1}(x)}{Q_n(x)} > 1. \qquad (7)$$

$$\frac{Q_n(x)}{Q'_n(x)} > 1. \tag{8}$$

These properties, however, have never rigorously proved mathematically and have remained unsolved for many years. In this paper, we try to give a rigorous proof of these properties. Some new refined results for $Q_n$ and $Q'_n$ have been obtained, which are listed below:

$$\frac{Q'_{n+1}(x)}{Q'_n(x)} > \frac{2n+3}{2n+1}. \tag{9}$$

$$\frac{Q_{n+1}(x)}{Q_n(x)} > \frac{2n+3}{2n+1}. \tag{10}$$

$$Q_{n+1}(x) = Q'_n(x) + x(n+1)\left|h^{(2)}_{n+1}(x)\right|^2. \tag{11}$$

$$Q'_{n+1}(x) = Q_n(x) + x(n+1)\left|h^{(2)}_n(x)\right|^2. \tag{12}$$

$$\frac{Q_{n+1}(x) - Q'_n(x)}{Q'_{n+1}(x) - Q_n(x)} > 1. \tag{13}$$

Equations (11) and (12) give the alternating recurrence relations for the modal quality factor $Q_n(x)$ and $Q'_n(x)$. We have also obtained the following new power series expansions for the modal quality factors

$$Q_n(x) = \sum_{m=0}^{n} \frac{q(n,m)}{x^{2(n-m)+1}}, \tag{14}$$

and

$$Q'_n(x) = \sum_{m=0}^{n} \frac{q'(n,m)}{x^{2(n-m)+1}}, \tag{15}$$

where the expansion coefficients are given by

$$q(n,m) = \frac{\left[n(n+2)+(n-m)^2-m\right](2n-m)!(2n-2m)!}{[(n-m)!]^2 (n+1-m)m! 2^{2(n-m)+1}}, \tag{16}$$

$$q'(n,m) = \frac{m(2n+1-m)(2n-m)!(2n-2m)!}{[(n-m)!]^2 (n+1-m)m! 2^{2(n-m)+1}}. \tag{17}$$

## II. PROPERTIES OF THE MODAL QUALITY FACTOR $Q'_n$

The property (4) will be established by using the method of mathematical induction. As the first step of the proof, we need to show that (4) is true for the case of $n=1$. This can be easily done from the definitions of spherical Bessel functions. Indeed we have $Q'_1(x) = 1/x > 0$. As the inductive step, we assume that (4) is true for an arbitrary $n$

$$Q'_n(x) > 0, \tag{18}$$

and we need to show that (4) holds for $n+1$. According to (2), we may write

$$Q'_{n+1}(x) = x - \frac{x^3}{2}\left[\left|h^{(2)}_{n+1}(x)\right|^2 - j_n(x)j_{n+2}(x) - n_n(x)n_{n+2}(x)\right]. \tag{19}$$

Making use of the recurrence relations of spherical Bessel functions, we may find that

$$j_n(x)j_{n+2}(x) + n_n(x)n_{n+2}(x) =$$
$$-\left|h^{(2)}_n(x)\right|^2 + \frac{2n+3}{2n+1}\left[\left|h^{(2)}_{n+1}(x)\right|^2 + j_{n-1}(x)j_{n+1}(x) + n_{n-1}(x)n_{n+1}(x)\right].$$

Thus we obtain

$$\left|h^{(2)}_{n+1}(x)\right|^2 - j_n(x)j_{n+2}(x) - n_n(x)n_{n+2}(x) =$$
$$-\frac{2}{2n+1}\left[\left|h^{(2)}_{n+1}(x)\right|^2 + \left|h^{(2)}_n(x)\right|^2\right]$$
$$+\frac{2n+3}{2n+1}\left[\left|h^{(2)}_n(x)\right|^2 - j_{n-1}(x)j_{n+1}(x) - n_{n-1}(x)n_{n+1}(x)\right].$$

Substituting this into (19) and rearranging terms, we have

$$Q'_{n+1}(x) =$$
$$\frac{2n+3}{2n+1}Q'_n(x) + \frac{x^3}{2n+1}\left[\left|h^{(2)}_{n+1}(x)\right|^2 + \left|h^{(2)}_n(x)\right|^2\right] - \frac{2}{2n+1}x. \tag{20}$$

This gives recurrence relation for the modal quality factor $Q'_n$. Utilizing the inequality (43) in the Appendix leads to

$$Q'_{n+1}(x) > \frac{2n+3}{2n+1}Q'_n(x) + \frac{x^3}{x^2}\frac{2}{2n+1} - \frac{2}{2n+1}x = \frac{2n+3}{2n+1}Q'_n(x). \tag{21}$$

We have thus proved the properties (5) and (9). From (21) and the assumption (18), we immediately obtain

$$Q'_{n+1}(x) > 0.$$

This completes the proof of the property (4).

A question may be raised whether there exists a constant $c$, greater than $(2n+3)/(2n+1)$, such that

$$Q'_{n+1}(x) > cQ'_n(x) \tag{22}$$

The answer is negative. In fact, we have

$$Q'_{n+1}(x) \to \frac{2n+3}{2n+1}Q'_n(x), \text{ as } x \to \infty$$

from (20) and the asymptotic behavior of spherical Hankel functions. Therefore the number $(2n+3)/(2n+1)$ is the maximum possible value of $c$ in (22).

### III. PROPERTIES OF MODAL QUALITY FACTOR $Q_n$ AND THE ALTERNATING RECURRENCE RELATIONS

Similarly the property (6) will be demonstrated by the method of mathematical induction. As the first step of the proof, we need to show that (6) is valid for $n=1$. From the definitions of spherical Bessel functions, we may easily obtain $Q_1(x) = x^{-1} + x^{-3} > 0$. Thus the property (6) has been established for the case of $n=1$. As the inductive step, we assume that (6) is true for an arbitrary $n$

$$Q_n(x) > 0. \quad (23)$$

We need to show that (6) also holds for the case of $n+1$. Let us consider

$$Q_{n+1}(x) = x - \left|h_{n+1}^{(2)}(x)\right|^2 \left[\frac{x^3}{2} + x(n+2)\right] - \frac{x^3}{2}\left|h_{n+2}^{(2)}(x)\right|^2 \quad (24)$$
$$+ \frac{x^2(2n+5)}{2}\left[j_{n+1}(x)j_{n+2}(x) + n_{n+1}(x)n_{n+2}(x)\right].$$

By use of the recurrence relations for spherical Bessel functions, we may easily find the following relations

$$\left|h_{n+2}^{(2)}(x)\right|^2 = \frac{(2n+3)^2}{x^2}\left|h_{n+1}^{(2)}(x)\right|^2 + \left|h_n^{(2)}(x)\right|^2$$
$$-\frac{2(2n+3)}{x}\left[j_{n+1}(x)j_n(x) + n_{n+1}(x)n_n(x)\right]$$

and

$$j_{n+1}(x)j_{n+2}(x) + n_{n+1}(x)n_{n+2}(x) =$$
$$\frac{2n+3}{x}\left|h_{n+1}^{(2)}(x)\right|^2 - j_n(x)j_{n+1}(x) - n_n(x)n_{n+1}(x).$$

Substituting these relations into (24) and ignoring the tedious derivations, we obtain

$$Q_{n+1}(x) = Q_n(x) + x(n+1)\left[\left|h_{n+1}^{(2)}(x)\right|^2 + \left|h_n^{(2)}(x)\right|^2\right] \quad (25)$$
$$-x^2\left[j_{n+1}(x)j_n(x) + n_{n+1}(x)n_n(x)\right].$$

Making use of (3),(4) and (25), we have

$$Q_{n+1}(x) = Q'_n(x) + x(n+1)\left|h_{n+1}^{(2)}(x)\right|^2 > 0.$$

This completes the proof of (6). Note that we have also proved the alternating recurrence relation (11).

According to (3), we may write

$$Q_{n+1} - Q'_{n+1} =$$
$$\left|h_{n+1}^{(2)}(x)\right|^2 (n+1)x - x^2\left[j_{n+1}(x)j_n(x) + n_{n+1}(x)n_n(x)\right].$$

Substituting this into (25), we immediately get the alternating recurrence relation (12):

$$Q'_{n+1}(x) = Q_n(x) + x(n+1)\left|h_n^{(2)}(x)\right|^2.$$

It follows from (9) that

$$Q'_n(x) > \frac{2n+1}{2n-1}Q'_{n-1}(x) > \frac{2n+3}{2n+1}Q'_{n-1}(x). \quad (26)$$

From the alternating recurrence relation (11), we may obtain

$$Q_n(x) = Q'_{n-1}(x) + xn\left|h_n^{(2)}(x)\right|^2. \quad (27)$$

Multiplying the above by $(2n+3)/(2n+1)$ and subtracting the resultant equation from (11) yield

$$Q_{n+1}(x) - \frac{2n+3}{2n+1}Q_n(x) = Q'_n(x) - \frac{2n+3}{2n+1}Q'_{n-1}(x)$$
$$+ x(n+1)\left|h_{n+1}^{(2)}(x)\right|^2 - x\left(n + \frac{2n}{2n+1}\right)\left|h_n^{(2)}(x)\right|^2. \quad (28)$$

Now making use of (9), (26) and (45) in the Appendix, we have

$$Q_{n+1}(x) - \frac{2n+3}{2n+1}Q_n(x) > 0. \quad (29)$$

Thus we have proved (10) and also (7). Similarly it can be shown that the number $(2n+3)/(2n+1)$ is the maximum possible value in (29).

Making use of (20), (28) can be expressed by

$$Q_{n+1}(x) - \frac{2n+3}{2n+1}Q_n(x) =$$
$$+ x(n+1)\left|h_{n+1}^{(2)}(x)\right|^2 - x\left(n + \frac{2n}{2n+1}\right)\left|h_n^{(2)}(x)\right|^2 \quad (30)$$
$$+ \frac{x^3}{2n+1}\left[\left|h_{n+1}^{(2)}(x)\right|^2 + \left|h_n^{(2)}(x)\right|^2\right] - \frac{2}{2n+1}x.$$

This is the recurrence relation for the modal quality factor $Q_n(x)$.

### IV. PROOF OF THE PROPERTY $Q_n(x) > Q'_n(x)$

To demonstrate the property (8), we have to resort to a useful property of the Lommel's polynomial [11]

$$J_{\nu+n+1}(x)J_{-\nu+n+1}(x) - J_{-\nu-n-1}(x)J_{\nu-n-1}(x)$$
$$= \frac{2}{\pi z}(-1)^n \sin\nu\pi R_{2n+1,\nu-n}(x), \quad (31)$$

where $J_\nu(x)$ stands for the Bessel functions, and $R_{2n+1,\nu-n}(x)$ is the Lommel's polynomial defined by

$$R_{2n+1,\nu-n}(x) =$$
$$\sum_{m=0}^{n} \frac{(-1)^m (2n+1-m)!\Gamma(\nu-n+2n+1-m)}{m!(2n+1-2m)!\Gamma(\nu-n+m)} \left(\frac{x}{2}\right)^{2(m-n)-1}.$$

if $\nu$ is not an integer. Here $\Gamma$ denotes the gamma function. For $\nu = 1/2$, (31) can be written as

$$x^2 \left[ j_{n+1}(x)j_n(x) + n_{n+1}(x)n_n(x) \right]$$
$$= 2\sum_{m=0}^{n} \frac{(2n+1-m)(2n-m)!(2n-2m)!}{[(n-m)!]^2 m!}(2x)^{2(m-n)-1}. \quad (32)$$

It follows from (42) in the Appendix that

$$x(n+1)\left|h_n^{(2)}(x)\right|^2 = 2\sum_{m=0}^{n} \frac{(n+1)(2n-2m)!(2n-m)!}{[(n-m)!]^2 m!}(2x)^{2(m-n)-1}. \quad (33)$$

From (3), (32) and (33) we obtain

$$Q_n(x) - Q'_n(x)$$
$$= 2\sum_{m=0}^{n} \frac{(n-m)(2n-m)!(2n-2m)!}{[(n-m)!]^2 m!}(2x)^{2(m-n)-1} > 0. \quad (34)$$

since $m \leq n$. The property (8) is thus validated.

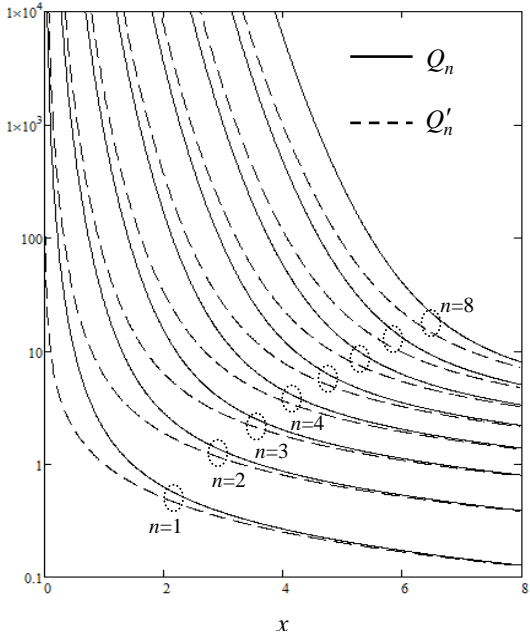

Figure 1 Modal quality factors

As a demonstration of the property (8), the modal quality factors $Q_n$ and $Q'_n$ are plotted in Figure 1. Note that both $Q_n$ and $Q'_n$ are monotonically decreasing functions of the variable $x = ka$. Physically this implies that the stored energy outside a sphere that encloses the antenna decreases rapidly as the radius of the sphere increases, and most of the stored energy exists in the near field region of the antenna. Also note that the difference between $Q_n$ and $Q'_n$ decreases as $ka$ increases. Physically this means that the stored electric energy will gradually approach to the stored magnetic energy outside the sphere as the radius $a$ increases.

V. AN APPLICATION

The properties (4) and (6) can be used to demonstrate that the stored energies around an antenna are always positive. The stored energy around an antenna is defined as the difference between the total energy and the radiated energy produced by the antenna:

$$\tilde{W}_e + \tilde{W}_m = W_e + W_m - \frac{r}{v}P^{rad}, \quad r \to \infty. \quad (35)$$

where $W_e$ and $\tilde{W}_e$ are the total and stored electric field energies around the antenna respectively; $W_m$ and $\tilde{W}_m$ are the total and stored magnetic field energies around the antenna respectively; $P^{rad}$ is the radiated power from the antenna; $r$ is the radius of the sphere that encloses the antenna, and $v = 1/\sqrt{\mu\varepsilon}$ is the speed of radiated wave in the space of medium parameters $\mu$ and $\varepsilon$. Note that the definition (35) is independent of the coordinates since the coordinate variable $r$ will eventually be sent to infinity.

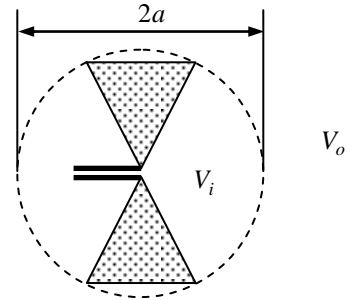

Figure 2 Antenna and the circumscribing sphere

Let $a$ be the radius of the circumscribing sphere of the antenna. The whole space can then be divided into the region $V_i$ enclosed by the circumscribing sphere and the region $V_o$ outside the circumscribing sphere, as illustrated in Figure 2. Accordingly, the total energy can be decomposed into the sum of two parts: the part contained in the circumscribing sphere and the part outside the circumscribing sphere. We use the subscript '$i$' and '$o$' to designate the corresponding parts of the stored energies. Then (35) can be written as

$$\tilde{W}_e + \tilde{W}_m = (W_e + W_m)_i + (W_e + W_m)_o - \frac{r}{v} P^{rad}, \quad r \to \infty. \quad (36)$$

The first term on the right-hand side denotes the total energy inside the circumscribing sphere and is positive and finite since the circumscribing sphere is finite. The last two terms denote the stored energy outside the circumscribing sphere and can be expressed as [4][6][9][10]

$$(W_e + W_m)_o - \frac{r}{v} P^{rad} = (\tilde{W}_e)_o + (\tilde{W}_m)_o, \quad (37)$$

where

$$(\tilde{W}_e)_o = \sum_{n=1}^{\infty}(a_n^2 Q_n' + b_n^2 Q_n), \quad (\tilde{W}_m)_o = \sum_{n=1}^{\infty}(a_n^2 Q_n + b_n^2 Q_n'),$$

with $a_n$ and $b_n$ being positive. Since both $Q_n$ and $Q_n'$ are positive, the stored energy outside the circumscribing sphere given by (37) is also positive. Thus the total stored energy defined by (36) is positive.

## VI. CONCLUDING REMARKS

We finally remark that the property (13) is an inference from the alternating recurrence relations (11) and (12). The new power series expansions (14) and (15) for the modal quality factors can be derived by using (32), (34) and (42) in the Appendix. The derivations are straightforward but rather tedious, and the details are omitted due to space limit. Note that the power series expansion (14) is much simpler than the one given by Collin and Rothschild [2]. For convenience, the expansion coefficients $q(n,m)$ and $q'(n,m)$ defined by (16) and (17) are listed in Table I and II for $n = 1-4$.

TABLE I VALUES OF $q(n,m)$

| n | $q(n,0)$ | $q(n,1)$ | $q(n,2)$ | $q(n,3)$ | $q(n,4)$ |
|---|---|---|---|---|---|
| 1 | 1 | 1 | | | |
| 2 | 18 | 6 | 3 | | |
| 3 | 675 | 135 | 21 | 6 | |
| 4 | $4.41 \times 10^4$ | $6.3 \times 10^3$ | 585 | 55 | 10 |

TABLE II VALUES OF $q'(n,m)$

| n | $q'(n,0)$ | $q'(n,1)$ | $q(n,2)'$ | $q'(n,3)$ | $q'(n,4)$ |
|---|---|---|---|---|---|
| 1 | 0 | 1 | | | |
| 2 | 0 | 3 | 3 | | |
| 3 | 0 | 45 | 15 | 6 | |
| 4 | 0 | $1.575 \times 10^3$ | 315 | 45 | 10 |

All the new results obtained in this paper have been validated numerically. It should be mentioned that the properties of modal quality factors are essentially the properties of the spherical Bessel functions, and therefore they are universally applicable. Their implications are surprisingly large and unrecognized. For example, the properties (4), (6) and (8) imply that

$$j_n(x)j_{n+1}(x) + n_n(x)n_{n+1}(x) > 0, \quad (38)$$

$$j_{n-1}(x)j_{n+1}(x) + n_{n-1}(x)n_{n+1}(x) > \left|h_n^{(2)}(x)\right|^2 - \frac{2}{x^2}, \quad (39)$$

$$\frac{n+1}{x}\left|h_n^{(2)}(x)\right|^2 < j_n(x)j_{n+1}(x) + n_n(x)n_{n+1}(x) < \frac{n+1}{x}\left|h_{n+1}^{(2)}(x)\right|^2. \quad (40)$$

for $x > 0$. These results are purely mathematical and are by no means trivial.

## APPENDIX

### PROOF OF THE TWO INEQUALITIES FOR HANKEL FUNCTIONS

Another useful property of the Lommel's polynomial is [11]

$$J_{\nu+n}(x)J_{n+1-\nu}(x) + J_{-\nu-n}(x)J_{-n-1+\nu}(x)$$
$$= \frac{2}{\pi x}(-1)^n \sin \nu \pi R_{2n,\nu-n}(x), \quad (41)$$

where $R_{2n,\nu-n}(x)$ is Lommel's polynomial defined by

$$R_{2n,\nu-n}(x) = \sum_{m=0}^{n} \frac{(-1)^m (2n-m)! \Gamma(\nu - n + 2n - m)}{m!(2n-2m)! \Gamma(\nu - n + m)} \left(\frac{x}{2}\right)^{-2n+2m},$$

if $\nu$ is not an integer. For $\nu = 1/2$, (41) can be written as

$$\left|h_n^{(2)}(x)\right|^2 = \frac{1}{x^2} \sum_{m=0}^{n} \frac{(2n-m)!(2n-2m)!}{[(n-m)!]^2 m!}(2x)^{2(m-n)}. \quad (42)$$

This can be rewritten as

$$\left|h_n^{(2)}(x)\right|^2 = \frac{1}{x^2}[1 + a(x)],$$

where $a(x)$ is positive for $x > 0$. This implies

$$\left|h_n^{(2)}(x)\right|^2 > \frac{1}{x^2}. \quad (43)$$

Note that

$$\left|h_{n-1}^{(2)}(x)\right|^2 = \frac{1}{x^2} \sum_{m=0}^{n-1} \frac{(2(n-1)-2m)!(2(n-1)-m)!}{[(n-1-m)!]^2 m!}(2x)^{2(m-n+1)}.$$

This can be rewritten as

$$\left|h_{n-1}^{(2)}(x)\right|^2 = \frac{1}{x^2} \sum_{m=1}^{n} \frac{(2n-2m)!(2n-m)!}{(2n/m-1)[(n-m)!]^2 m!}(2x)^{2(m-n)}.$$

Since $(2n/m - 1) \geq 1$ for $1 \leq m \leq n$, we have

$$\begin{aligned}\left|h_{n-1}^{(2)}(x)\right|^2 &= \frac{1}{x^2}\sum_{m=1}^{n}\frac{(2n-2m)!(2n-m)!}{(2n/m-1)[(n-m)!]^2 m!}(2x)^{2(m-n)} \\ &\leq \frac{1}{x^2}\sum_{m=1}^{n}\frac{(2n-2m)!(2n-m)!}{[(n-m)!]^2 m!}(2x)^{2(m-n)} \qquad (44)\\ &= \left|h_n^{(2)}(x)\right|^2 - \left[\frac{(2n)!}{n!}\right]^2 (2x)^{-2n}.\end{aligned}$$

This implies

$$\left|h_n^{(2)}(x)\right|^2 > \left|h_{n-1}^{(2)}(x)\right|^2. \qquad (45)$$